\documentclass[english,preprint,onecolumn,showpacs,preprintnumbers,amsmath,amssymb,showkeys]{revtex4-2}

\usepackage{chemformula} 
\usepackage[T1]{fontenc} 
\usepackage{bm}
\usepackage[normalem]{ulem}
\usepackage[T1]{fontenc}
\usepackage{color}
\usepackage{units}
\usepackage{amssymb}
\usepackage{graphicx}
\usepackage{esint}
\usepackage{bm}
\usepackage{natbib}
\usepackage{ulem}
\usepackage{babel}
\usepackage[colorlinks=true, citecolor=red]{hyperref}

\begin{document}

\title[Graphene photo-doping]{Optical control of multiple resistance levels in graphene for memristic applications  }

\author{Harsimran Kaur Mann$^1$, Mainak Mondal$^1$, Vivek Sah$^1$,  Kenji Watanabe$^2$, Takashi Taniguchi$^3$, Akshay Singh$^1$}
\email{aksy@iisc.ac.in}
\author{Aveek Bid$^1$}
\email{aveek@iisc.ac.in}
\affiliation{$^1$Department of Physics, Indian Institute of Science, Bangalore 560012, India \\
$^2$ Research Center for Electronic and Optical Materials, National Institute for Materials Science, 1-1 Namiki, Tsukuba 305-0044, Japan\\
$^3$ Research Center for Materials Nanoarchitectonics, National Institute for Materials Science,  1-1 Namiki, Tsukuba 305-0044, Japan\\}

\begin{abstract}

Neuromorphic computing has emphasized the need for memristors with non-volatile, multiple conductance levels. This paper demonstrates the potential of hexagonal boron nitride (hBN)/graphene heterostructures to act as memristors with multiple resistance states that can be optically tuned using visible light. The number of resistance levels in graphene can be controlled by modulating doping levels, achieved by varying the electric field strength or adjusting the duration of optical illumination. Our measurements show that this photodoping of graphene results from the optical excitation of charge carriers from the nitrogen-vacancy levels of hBN to its conduction band, with these carriers then being transferred to graphene by the gate-induced electric field. We develop a quantitative model to describe our observations. Additionally, utilizing our device architecture, we propose a memristive crossbar array for vector-matrix multiplications.
\end{abstract}

\maketitle

\section{Introduction}
Neuromorphic computing (NC), inspired by computing in our brain, has emerged as a new paradigm beyond von Neumann computing. NC provides a low-energy alternative to traditional von Neumann architectures, promoting sustainable computation in our modern information age. The critical hardware building blocks for NC are memristors (artificial synapse), neural processing units, and threshold switches (artificial neurons) ~\cite{sangwan_neuromorphic_2020,walters_review_2023,lu_ferroelectric_2014,wu_high_2020,yan_moire_2023}. A memristor is a fundamental electronic component where the resistance of the channel is dependent on the charge that has flown through it; it is a resistor with a memory. This charge-history dependence of resistance can be exploited in machine learning frameworks where computation can be carried out via cross-bar arrays, and synaptic weights (such as weights for neural networks) can be modified by a certain number of electrical pulses, leading to in-memory computing. A low-power, non-volatile memristor is essential to exploit NC's benefits fully.

A prerequisite for memristic action is a platform with non-volatile doping.
Chemical and electrostatic doping are the two most commonly used techniques to induce charge carriers in 2-D channels. Chemical doping is done by hetero-atom substitution and adsorption of molecular adsorbates onto graphene and other 2D materials. There are several drawbacks of this approach, including limited control over doping concentration, introduction of uncontrolled structural defects and lattice strains, unintentional impurity introduction, challenges in maintaining stability and diffusion control, and the sensitivity of the sample to processing conditions.~\cite{liu_chemical_2011,wehling_molecular_2008,jung_charge_2009,bruna_observation_2010,zhan_fecl3-based_2010,zhao_intercalation_2011,zhao_charge_2010,singh_molecular_2012,medina_tuning_2011}.

 Electrostatic doping involves doping using an external local gate. While this technique avoids most of the disadvantages of chemical doping listed above, the magnitude of doping attainable is limited by the breakdown voltage of the gate dielectric.~\cite{ryu_atmospheric_2010,dean_boron_2010,das_monitoring_2008,yan_electric_2007}. Liquid ionic gating has the potential of reaching higher doping levels than is achievable by dielectric gates but at the cost of device instability and non-scalability.~\cite{chen_electrochemical_2009,uesugi_electric_2013}

A viable alternative is optical doping of graphene on hexagonal boron nitride (hBN) substrates~\cite{tiberj_reversible_2013,aftab_programmable_2022,doi:10.1021/acsami.6b01727,ju_photoinduced_2014,roy_graphenemos2_2013,doi:10.1021/nn402354j,song_deep-ultraviolet_2021,doi:10.1021/acs.nanolett.5b04441,kim_multilevel_2019,liu_graphene_2014,xiang_two-dimensional_2018,tran_two-terminal_2019,lee_photoinduced_2020,gorecki_optical_2018,seo_direct_2014,miller_nonvolatile_2020}. hBN has a band gap of 6~eV with several intermediate defect states that can be optically excited~\cite{attaccalite_coupling_2011,weston_native_2018,PhysRevB.97.064101,strand_properties_2020}. This method involves the controlled optical excitation of charge carriers from defect states of hBN and their transfer to graphene using an external electric field. This technique is reversible without adversely affecting transport mobility and defect density.

In this article, we present an in-depth study of optical doping in hBN/graphene heterostructures using visible light. Our research reveals that electrons in the nitrogen-vacancy defect state in hBN can be optically excited with violet light and transferred to graphene through electrostatic gating. This technique enables high-density doping of graphene, with the doping level controlled by gate voltage or illumination time. The dynamics of this doping process, measured with millisecond temporal resolution, can be adjusted over three orders of magnitude by varying the illumination wavelength and optical power. Furthermore, we demonstrate that a graphene/hBN heterostructure can function as a memristor, exhibiting multiple resistance levels. The device's compact structure and room-temperature operation enhance its potential for scalable, room-temperature neuromorphic applications.

\section{Results}

\begin{figure}[t]
\centering
   \includegraphics[width=\columnwidth]{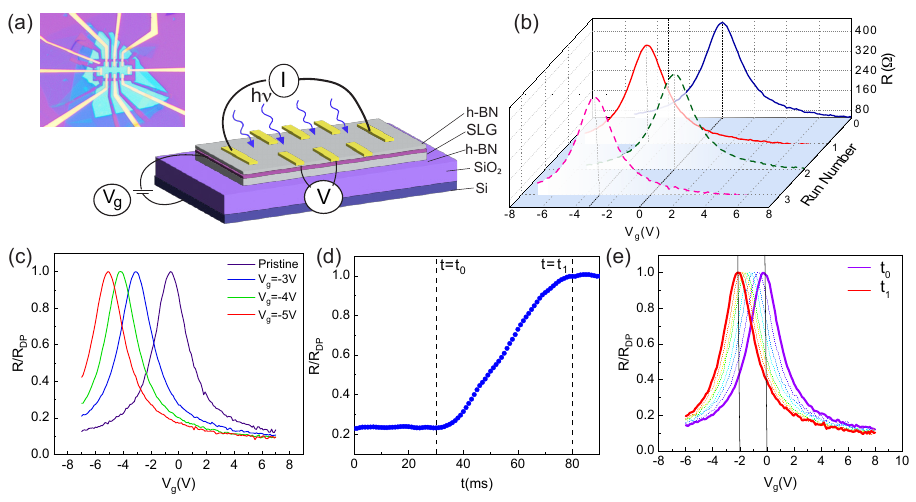}
  \caption{\textbf{Device characteristics and optical doping:} (a) Optical image of the device and a schematic of the device structure. (b) Longitudinal resistance as a function of gate voltage for the pristine sample before illumination (blue solid line), after photodoping at $V_g^*=-4$~V (red solid line), after erasing the doping (green dotted line), and after photodoping again at $V_g^*=-4$~V (pink dotted line). (c) Representative plots of deterministic shifting of the Dirac point to desired values of $V_g$ by photodoping. $R_{DP}$ is the four-probe resistance at the Dirac point. (d) Change in the longitudinal $R$ as a function of time for constant power and gate voltage on exposure to light. $t=t_0$ marks when the light was turned on and $t=t_1$ when the light was turned off. (e) Snapshots of $R-V_g$ plots as the Dirac point progressively shifts from $V_g = 0$~V at $t=t_0$ (solid purple line) to $V_g* = -2$~V at $t=t_1$ (solid red line).}
  \label{fig1}
   \end{figure}

Single-layer graphene devices encapsulated between thin hBN crystals were fabricated using the dry transfer method (Supporting Information, section S1). 1-D electrical contacts to the graphene were achieved by lithography and dry etching, followed by Cr/Au metallization (Fig.~\ref{fig1}(a)). A back-gate voltage, $V_g$, tuned the charge carrier number density. Electrical transport measurements were performed using a low-frequency AC measurement technique. The Dirac point (maxima in the device resistance $R$) is attained at $V_g=0$~V (Fig.~\ref{fig1}(b), solid blue line), attesting to the absence of charged impurities in the graphene channel.

The optoelectronic measurements were carried out at room temperature, and the sample was illuminated by using either an LED of wavelength, $\lambda=427$~nm or a Ti Sapphire pulsed laser (80~MHz repetition rate, $100$~fs pulse width). To photo-dope the device, we use the following protocol: The gate voltage is set to a desired value $V_g^*$, and the device is exposed to the light of wavelength 427~nm  of intensity $I=62~\mathrm{W/ m^2}$, till the resistance saturates to the value of resistance at Dirac point. The light is then turned off, and the gate response of the device is measured. It was found that the entire $R-V_g$ plot shifts with the Dirac point at $V_g^*$ (Fig.~\ref{fig1}(b) -- solid red line; in this example
 $V_g^* = -4$~V), establishing that the device is now electron-doped. We refer to this step as the `SET' protocol wherein the Dirac point can be set deterministically at any desired value of $V_g$ (Fig.~\ref{fig1}(c)). To `RESET,' the device is exposed to a higher light intensity $I=458~\mathrm{W/ m^2}$ at $V_g^*=0$~V until the Dirac point shifts to $V_g = 0$~V. This protocol brings the Dirac point of the device back to $V_g= 0$~V (Fig.~\ref{fig1}(b) -- dotted green line). As illustrated in Fig.~\ref{fig1}(b), the process can be repeated without degradation in the device characteristics.

Fig.~\ref{fig1}(d) shows the time dependence of the device's normalized longitudinal resistance $R/R_D$ during the `SET' protocol with $V_g^*=-2$~V using LED of $\lambda=427~$~nm and $I=62~\mathrm{W/ m^2}$. Here, $R_D$ is the resistance at the Dirac point. Upon illumination, the device resistance, $R$, increases rapidly with time, saturating in about 50 milliseconds to $R = R_D$. Exposure for a longer duration had no discernible effect on the channel resistance. Notably, upon turning off the illumination, the device's resistance remains unchanged.

Fig.~\ref{fig1}(e) shows snapshots of the $R-V_g$ data in 10-second intervals. During this measurement, the light was turned on for 10 seconds with $V_g^*$ set at $-2$~V. The illumination was turned off, and the $R-V_g$ curve was measured. The process is repeated multiple times to generate the plots in Fig.~\ref{fig1}(e). The intensity of the light was kept very low (I=$4~\mathrm{W/m^2}$) to allow for a much slower rate of resistance change (for easier observations). One can see that the entire transfer curve shifts gradually to the left until the Dirac point reaches $V_g^* = -2$~V. These measurements establish that the Dirac point can be moved deterministically to any value of the gate voltage either by controlling the exposure time (Fig.\ref{fig1}(e)) or the value of $V_g^*$ at which the device is illuminated (Fig.~\ref{fig1}(c)).

\begin{figure*}[t]
\centering
\includegraphics[width=\columnwidth]{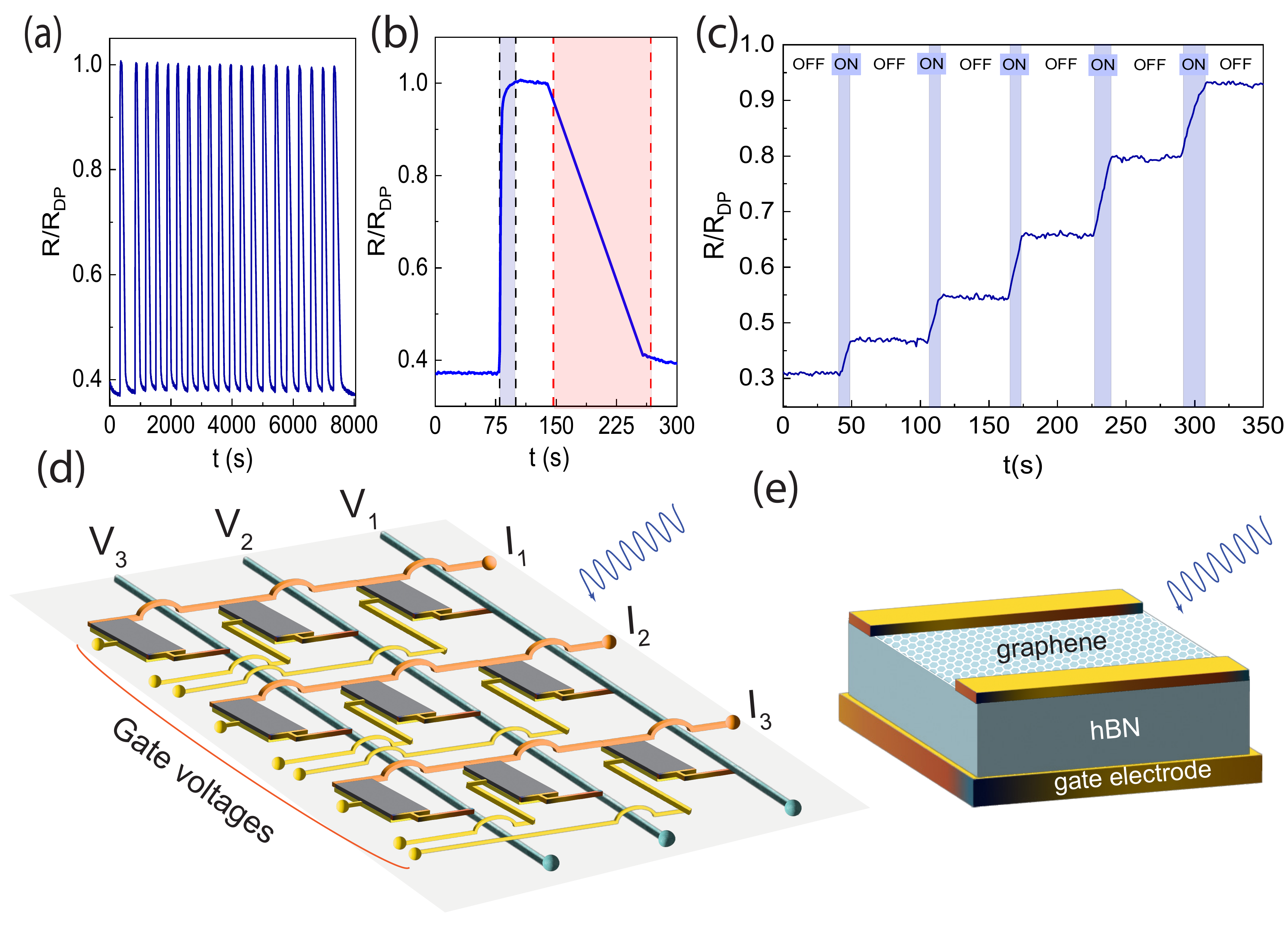}
  \caption{\textbf{Reproducibility and multiple states:} (a) Optical switching of the device's resistance between two well-defined values at $V_g^*=-5V$ with led light source of $\lambda=427$~nm and I=$62~\mathrm{W/m^2}$  (b) Resistance versus time for a single light pulse. The `SET' time is significantly shorter than the `RESET' time. (c) Figure illustrating optical pinning of $R$ to multiple values. The blue shaded region marks when the light was switched on (hence  $R$ increased with $t$). The white regions show the time the light was off. (d) Schematic of the proposed cross bar geometric memristor. (e) Schematic of the individual device forming a cell of the memristor. }
 \label{fig2}
 \end{figure*}

After setting up the SET/RESET protocol, we now show that the resistance of the sample at a specific gate voltage can be repeatedly alternated between two distinct values (Fig.~\ref{fig2}(a)). The data for a single light pulse is shown in Fig.~\ref{fig2}(b); we find that the `RESET' time is significantly larger than the `SET' time for electron doping. Below, we explain this observation.   Moreover, adjusting the exposure time makes switching between multiple resistance values possible, as depicted in Fig.~\ref{fig2}(c). In this measurement, the light was turned on for $10$~seconds (gray shaded region of the timeline), during which $R$ increased. The light was then turned off. The value of $R$ was stable at the value it reached when the illumination was cut off. This process can be repeated to produce multiple stable resistance levels in graphene.

\section{Application as memristor}

As shown in Fig.~\ref{fig2}(c), our device has at least six stable resistance states, with more resistance states also accessible by lower optical powers. Such an hBN/graphene heterostructure with multiple stable resistance values holds significant potential for developing memristor devices for vector-matrix multiplication and machine learning (ML) applications. ~\cite{ducry_ab_2022,xie_hexagonal_2022,maier_electro-photo-sensitive_2016,schranghamer_graphene_2020}. Several such devices can be fabricated in a cross-bar array for a typical linear algebra calculation, with voltages as inputs and currents as outputs. A new vector-matrix multiplication operation can be carried out by modifying the weights of each cross-bar intersection, i.e., by changing the resistance of the channels. For hardware implementation of ML training, the synaptic weights of a neural layer (each layer will be a separate cross-bar array) can be similarly modified.

We propose a cross-bar geometry schematically shown in Fig.~\ref{fig2}(d) to achieve the above objectives. Its compact footprint offers distinct advantages over other structures. Each device unit (shown schematically in Fig.~\ref{fig2}(d)) is individually gated; this architecture is easily achievable using modern nano-fabrication processes. Before each operation, the channel resistances of each device are initialized by a global incident optical beam and the application of distinct back-gate voltages to different devices. This process will `SET` the channel resistance of each device. Conversely, the `RESET` can be done by setting the desired gate voltages to zero and illuminating with a global incident optical beam.

\section{Origin of the phenomenon}

Photo-doping of the graphene channel requires charge transfer from hBN to graphene. The energy corresponding to violet light ($\lambda=427$~nm) is 2.9~eV, much lower than the band gap of hBN ($\approx 6$~eV), precluding photo-excitation of carriers from the valence band of hBN. We also find that the graphene channel remained undoped in a device with only the top hBN flake and without the bottom hBN flake upon using the same protocol described above  (Supporting Information, section S3). This study confirms that only the bottom hBN was responsible for the photodoping effect. Based on these observations, we sketch out a possible scenario below that explains all our experimental findings.

 \begin{figure*}
   \centering
\includegraphics[width=\columnwidth]{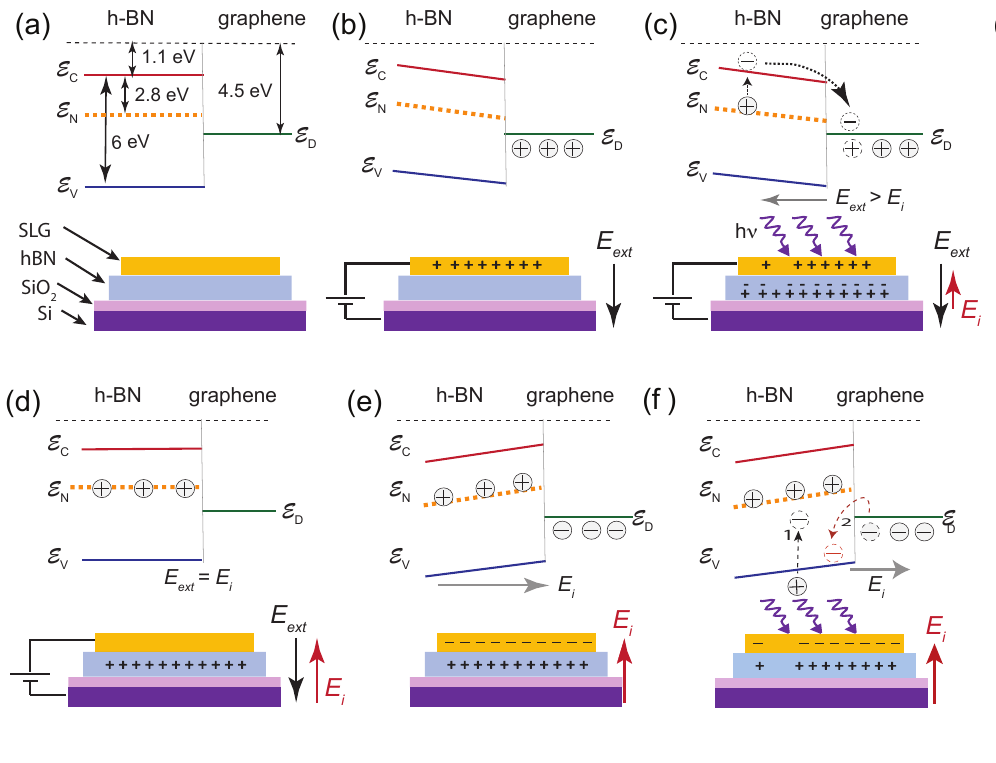}
  \caption{\textbf{Schematic energy band alignment (top panels) and charge distribution (bottom panels) in the device under different conditions. $\mathcal{E}_{C}$, $\mathcal{E}_{N}$ and $\mathcal{E}_{V}$ represent the conduction band edge, N-vacancy mid-gap state and valence band edge of hBN, respectively. }(a) For $V_g^*=0, h\nu =0$, hBN and graphene are charge neutral. (b) The graphene channel acquires a positive charge for $V_g^*< 0$ and $h\nu =0$. The resultant electric field $E_{ext}$ bends the bands of the hBN. (c) SET: For $V_g^*< 0$ and $h\nu \neq 0$, electrons are transferred from $\mathcal{E}_{N}$ (orange dotted line) to $\mathcal{E}_{C}$ (solid red line) of hBN. These electrons drift into graphene (represented by the dotted electrons) and reduce its net positive charge. (d) The doping process stops when $E_i = E_{ext}$. The hBN bands flatten out, graphene is charge neutral, and the hBN is left positively charged. (e) After the gate voltage is set to zero, graphene draws a net negative charge from the contacts, making it electron-doped and the total system charge neutral. (f) RESET: For $V_g^*=0$~V and $h\nu \neq 0$, electrons are excited from $\mathcal{E}_{v}$ (blue solid line) of the hBN to recombine with the holes in $\mathcal{E}_{N}$ (orange dotted line). The electrons from graphene (represented by the red dotted circles) drift into hBN $\mathcal{E}_{v}$ and recombine with the holes left behind.}
\label{fig3}
 \end{figure*}

Fig.~\ref{fig3}(a) is a schematic of the energy alignment of the bottom hBN and the graphene without photo-excitation and at $V_g=0$. Several mid-gap states in hBN can act as electron donors. Of these, the one most relevant for us is the defect state of nitrogen vacancies (marked as $\mathcal{E}_{N}$ in Fig.~\ref{fig3}). This level can have stable charge states of $+1$ or $0$~\cite{weston_native_2018}. A negative gate-voltage $V_g = V_g^*$ dopes the graphene channel with holes and creates an electric field $E_{ext}$  directed from graphene into the hBN (Fig.~\ref{fig3}(b)) leading to band bending. Illuminating the device with violet light excites electrons from $\mathcal{E}_{N}$ to the conduction band $\mathcal{E}_{C}$ of hBN. These electrons are funneled to the graphene channel under the influence of the electric field. The holes left at $\mathcal{E}_{N}$ in the hBN  generate an electric field $E_i$ in the direction opposite $E_{ext}$ (Fig.~\ref{fig3}(c)). The electron transfer process continues until the net electric field between graphene and hBN $E_{net} = E_{ext}-E_i$ becomes zero. Simple electrostatics arguments show that the effective number density in graphene becomes zero at this point~(Fig.~\ref{fig3}(d), which manifests as a shift of the charge neutrality point to $V_g^*$, (Fig.~\ref{fig3}(d)) and constitute the 'SET' protocol.

On reducing $V_g$ to zero, graphene draws negative charges from the metal contacts, making the net device charge neutral (Fig.~\ref{fig3}(e)). This charge configuration generates an electric field $E_i$ directed from the hBN to graphene. The consequent band bending and the fact that $\mathcal{E}_N>\mathcal{E}_D$ forbids electron transfer back from graphene to the hBN; graphene remains negatively charged, and hBN is positively charged (Fig.~\ref{fig3}(e)). This energy barrier to back-transfer electrons from graphene to hBN explains the long-term charge retention in graphene after the photodoping is completed.

\begin{figure*}
   \centering
\includegraphics[width=\columnwidth]{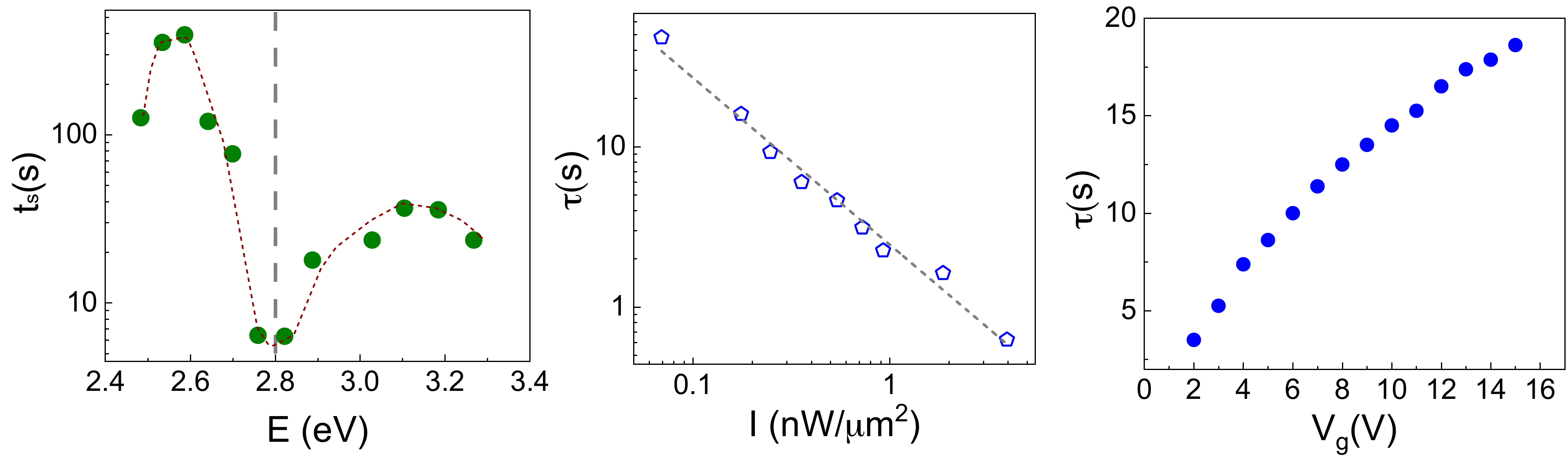}
  \caption{\textbf{Wavelength, Power and Gate Dependence of Doping Dynamics: }(a) Plot of time taken for the resistance to saturate, $t_s$ versus the laser photon energy. The data were taken at a fixed gate voltage $V_g^*=-3$~V and $P=70$~nW. The dotted red line is a guide to the eye. The dashed grey line marks $\mathcal{E}_N$. (b) Log-log plot of time constant $\tau$ versus the intensity of incident light. The data were taken at $V_g^* =-5$~V and $\lambda=427$~nm using LED. The open symbols are the measured data points, and the dashed grey line is a linear fit to the data. (c) Plot of $\tau$ versus the $V_g^*$. The data was taken for $\lambda=427$~nm and $I=62~\mathrm{W/m^2}$ using LED.}
 \label{fig4}
 \end{figure*}

Applying a $V_g^*=0$V with simultaneous exposure to light erased the doping, bringing the graphene's Dirac point back to $V_g=0$~V. Excess electrons need to be removed from graphene and transferred back to hBN for this to happen. Note, however, that this process seems energetically unfavorable as $\mathcal{E}_N > \mathcal{E}_D$ at the interface. The physical mechanism leading to this charge-neutralization of graphene is unclear. We propose a phenomenological scenario in which this `RESET' process is a two-step process involving (1) the transfer of electrons from the valence band of hBN to its mid-gap states due to optical excitation and (2) subsequent electron transfer from graphene to the empty valence band states of hBN due to electric field (Fig.~\ref{fig3}(f)). Consequently, the `RESET' process (involving electron transfer from graphene to hBN) is much slower than the `SET' process for electron doping. Device-level simulations are required to verify if the above scenario correctly captures the doping erasure process.

Next, we used a tunable pulse laser to study the wavelength dependence of the doping time $t_s$, which we define as the time taken for the resistance to saturate to $R=R_D$ on exposure to light at a fixed $V_g^* $. Fig.~\ref{fig4}(a) plots $t_s$ versus the laser photon energy for a constant laser power $P=70$~nW and $V_g^* = -3$~V. It shows a minimum in $t_s$ around $E=2.8$~eV. This photon energy corresponds to the optical absorption by valence nitrogen defect in hBN `$\mathcal{E}_N$,' leading credence to our understanding of the photodoping process ~\cite{attaccalite_coupling_2011}. The intensity and gate-dependent measurements were done using an LED source ($\lambda=427$~nm), and the time constant $\tau$ was extracted by fitting number density vs. time curve to equation (S1)(Fig.~S1). Intensity-dependent measurements at a fixed $V_g^*$  showed that $\tau$ is inversely proportional to the light intensity (Fig.~\ref{fig4}(b)). This dependence is understandable, as an increase in the intensity of photons leads to more free charges being produced in hBN, reducing the time taken to dope (see Supplementary Information for detailed derivation). For measurements performed at a fixed intensity of light, the time constant to dope should increase with an increase in the magnitude of $V_g^*$ (equivalently, of $E_{ext}$);  measurements confirm this (Fig.~\ref{fig4}(c)) (See Supplementary Information for a derivation).

\section{Conclusions}
To summarize, we demonstrate a reversible control of the Dirac point in the graphene-hBN heterostructure to encode the resistance values via a combined optical-electrical stimulus. The device's resistance can be modified by varying the gate voltage in the presence of an optical incident power or by fixing the gate voltage and illuminating it with multiple optical pulses. The switching time can be tuned by the incidence light wavelength (Fig.~\ref{fig4}(a)), light intensity (Fig.~\ref{fig4}(b)), or the gate voltage (Fig.~\ref{fig4}(c)) providing tremendous tunability of the properties of the device. The time taken to electron dope is much less compared to hole doping, and further experiments need to be done to make them comparable. It should be possible to control the switching time by modifying the defect density in hBN using electron irradiation or annealing processes. The ability to photoelectrically `SET,' `READ,' and `RESET' multiple stable and non-volatile resistance states of the device makes it ideal for use as a memristor.

\section*{Methods}
\subsection*{Device Fabrication}
\noindent The devices were fabricated using the dry transfer technique~\cite{tiwari_observation_2023}. Single-layer graphene (SLG) and hexagonal boron nitride (hBN) flakes were mechanically exfoliated onto a \ch{Si}/\ch{SiO2} substrates. The hBN flakes had a thickness of 25-30 nm. Electron beam lithography was used to define electrical contacts. This was followed by etching with a mixture of $\ch{CHF_3}$ (40 sscm) and $\ch{O_2}$ (10 sscm). The metallization was done with Cr/Au (5~nm/60 nm) to form the 1D electrical contacts with SLG.

\subsection*{Measurements}

\noindent All electrical transport measurements were performed at room temperature using a low-frequency AC measurement technique. For low-temperature measurements, the sample was cooled down in a cryostat to 4.7~K. For optoelectronic measurements, the sample was illuminated using either an LED of wavelength, $\lambda=427$~nm, or a Ti Sapphire pulsed laser (80~MHz repetition rate, $100$~fs pulse width).

\noindent \textbf{Acknowledgement}

\noindent A.B. acknowledges funding from the Department of Science \& Technology FIST program and the U.S. Army DEVCOM Indo-Pacific (Project number: FA5209   22P0166). K.W. and T.T. acknowledge support from JSPS KAKENHI (Grant Numbers 19H05790, 20H00354, and 21H05233). A.S. acknowledges funding from Indian Institute of Science start-up grant, DST Nanomission CONCEPT (Consortium for Collective and Engineered Phenomena in Topology) grant and project MoE-STARS-2/2023-0265. M.M. acknowledges Prime Minister’s Research Fellowship (PMRF).   \\

\noindent \textbf{Data availability}

\noindent The authors declare that the data supporting the findings of this study are available within the main text and its supplementary Information. Other relevant data are available from the corresponding author upon reasonable request.\\

\noindent \textbf{Author Contributions}

\noindent H.K.M, M.M., V.S., A.S., and A.B. conceptualized the study, performed the measurements, and analyzed the data. K.W. and T.T. grew the hBN single crystals. All the authors contributed to preparing the manuscript.\\

\noindent  \textbf{Competing interests:}

\noindent The authors declare no Competing Financial or Non-Financial Interests.\\

 \noindent \textbf{Supporting Information}

 \noindent Supporting information contains detailed discussions of the (S1) device fabrication, (S2)dependence of $\tau$ on $I$ and $V_g^*$, (S3) top or bottom hBN responsible for doping, (S4) doping at low temperatures and (S5) minimum detectable power.

\clearpage

\section*{Supplementary Materials}

	\renewcommand{\theequation}{S\arabic{equation}}
	\renewcommand{\thesection}{S\arabic{section}}
	\renewcommand{\thefigure}{S\arabic{figure}}
	\renewcommand{\thetable}{S\arabic{table}}
	\setcounter{table}{0}
	\setcounter{figure}{0}
	\setcounter{equation}{0}
	\setcounter{section}{0}

\section{\textbf{Device Fabrication}}

The devices were fabricated using the dry transfer technique~\cite{tiwari_observation_2023}. Single-layer graphene (SLG) and hexagonal boron nitride (hBN) flakes were mechanically exfoliated onto a \ch{Si}/\ch{SiO2} substrates. The hBN flakes had a thickness of 25-30 nm. The SLG flakes were identified from optical contrast under a microscope and later confirmed with Raman spectra. Electron beam lithography was used to define electrical contacts. This was followed by etching with a mixture of $\ch{CHF_3}$ (40 sscm) and $\ch{O_2}$ (10 sscm). The metallization was done with Cr/Au (5~nm/60 nm) to form the 1D electrical contacts with SLG.

    \begin{figure}[h]
\centering
   \includegraphics[width=0.5\columnwidth]{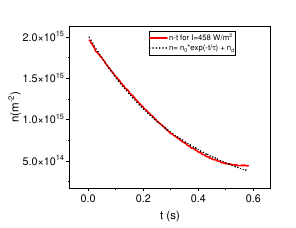}
  \caption{Plot of $n$ versus $t$; the data were taken at $I=458 \mathrm{W/m^2}$. The solid red line fits the data to the equation $n\left(t\right)=n_0e^{-t/\tau} + n_d$.}
   \label{fig_s1}
   \end{figure}

\begin{figure}[h]
\centering
   \includegraphics[width=\columnwidth]{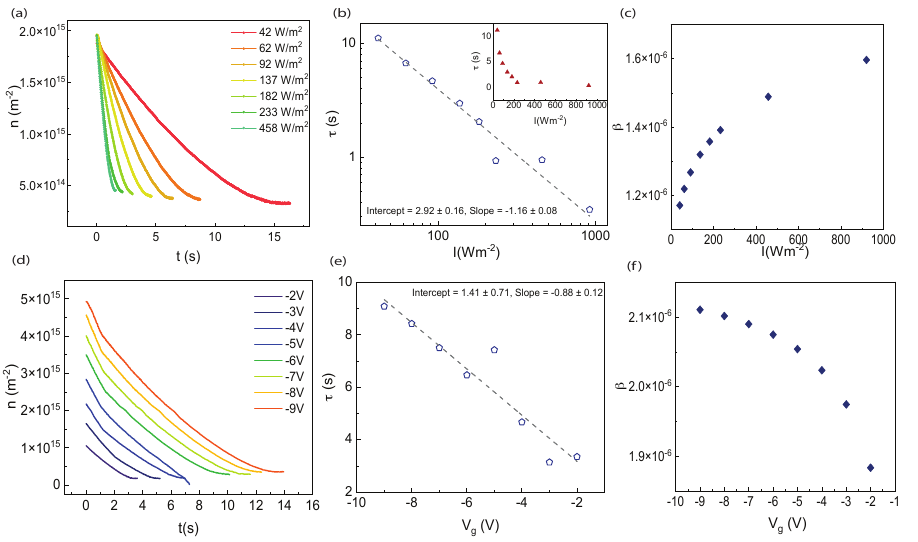}
  \caption{(a) $n-t$ curves for different values of the photon intensity $I$. The measurements were done for  $V_g^*=-5$~V. (b) Plot of $\tau$ versus $I$ on a log-log plot; the dotted line is a linear fit with slope $-1.16 \pm 0.08$. The inset shows the data on a linear plot. (c) Plot of $\beta$ for different values of $I$. (d) $n-t$ curves measured for different values of $V_g^*$. (e) Plot of $\tau$ versus $V_g^*$.  The dotted line is a linear fit with slope $-0.88 \pm 0.12$. (f) Plot of $\beta$ versus $V_g^*$.}
  \label{fig_s2}
   \end{figure}

\section{\textbf{Dependence of $\tau$ on $I$ and $V_g^*$}}

As discussed in the main manuscript, illuminating the device with the gate voltage held at a gate voltage $V_g = V_g^*$ leads to the shift of the Dirac point to $ V_g^*$. Consider the graphene channel with the gate voltage set to $ V_g^*$ before turning the light on. The channel is hole-doped to a value $n_0 = C_gV_g^*/e$. Here, $C_g$ is the capacitance of the bottom hBN layer.

On turning on the light, the hole density $n$ decreases exponentially with time with a time constant $\tau$ (Fig.~\ref{fig_s1}):
\begin{equation}
 n\left(t\right)=n_0e^{-t/\tau} + n_d
  \label{eq_1}
\end{equation}
where $n_d$ is the residual charge density at the Dirac point.
From Eqn.~\ref{eq_1}, we get:
\begin{equation}
\delta n\left(t\right)=n\left(t\right)-n\left(0\right)
            =n_0\left(e^{-t/\tau}-1\right) \label{eqn:deltan}
\end{equation}
It follows that the areal number density of electrons excited in time $t$ from the N-vacancy mid-gap state to the conduction band of hBN is:
\begin{eqnarray}
   \delta n(t)= I\beta t \label{Eqn:beta}
\end{eqnarray}
Here $I$ is the
number of photons incident on the device per unit area per unit time, $\beta$ is the efficiency of the electron excitation process in hBN. For simplicity, we assume that all the electrons produced in hBN are instantly swept to graphene by the process explained in the main manuscript. Combining Eqn.~\ref{eqn:deltan} and Eqn.~\ref{Eqn:beta}, we get
$$n_{0\ }{(e}^{-t/\tau}-1)=I\beta t$$
For $t/\tau<<1$,this reduces to
$${-n}_0t/\tau=I\beta t$$
or,
\begin{equation}
\tau={-C}_gV_g^*/eI\beta
\label{eq_4}
\end{equation}

Eqn.~\eqref{eq_4} implies that the time constant for doping should be directly proportional to $V_g^*$ and inversely proportional to $I$. Below, we probe the $I$ and $V_g^*$ dependence of $\tau$. We also get the values of $\beta$.

\paragraph{Intensity dependence:}

For these measurements, the gate voltage is kept constant at $V_g^*=-5$~V, and data are collected for different values of $I$. Some representative plots are shown in Fig.~\ref{fig_s2}(a). The data are fitted using Eqn.~\eqref{eq_1} to extract the values of $n_d$, $n_0$ and $\tau$. Fig.~\ref{fig_s2}(b) is the plot of $\tau$ versus $I$. A double log fit yields $\tau=10^{23.4} / I$ establishing $\tau$ to be inversely proportional to $I$ and matching the prediction of Eqn.~\eqref{eq_4}. From the fit, we get $\beta$ in the range $1.1\times10^{-6} - 1.6\times10^{-6}$, the data are plotted in  Fig.~\ref{fig_s2}(c).

\paragraph{Gate Voltage dependence:}

For this set of measurements, the light intensity is kept fixed at $I = 137$~$\mathrm{W/m^2}$. Data are collected for different values of $V_g^*$ (Fig.~\ref{fig_s2}(d)). The data are fitted using Eqn.~\eqref{eq_1} to extract the values of  $\tau$ (Fig.~\ref{fig_s2}(e)). The dotted line is a linear fit to the data. The excellent fit establishes that $\tau \propto -V_g^*$, matching the prediction of  Eqn.~\eqref{eq_4}. The slope of the curve ($\tau/V_g^*=-C_g/(eI\beta)$) yields $\beta$ to lie in the range of $1.9\times10^{-6} - 2.1\times10^{-6}$ (Fig.~\ref{fig_s2}(f)).

Given the approximations made in deriving Eqn~\eqref{eq_4}, its experimental verification, as shown in  Fig.~\ref{fig_s2}, is remarkable.

\section{\textbf{Top or bottom hBN responsible for doping}}
\begin{figure}[h]
\centering
   \includegraphics[width=\columnwidth]{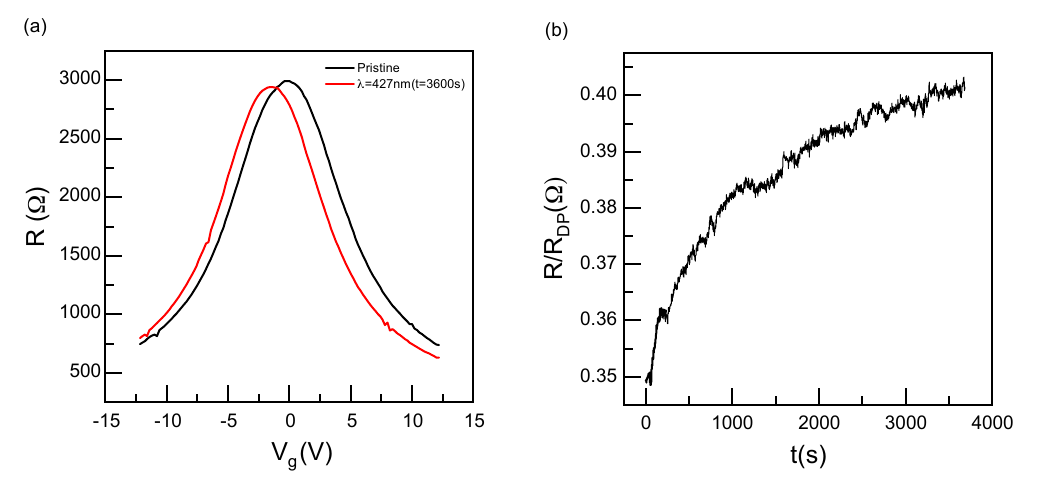}
  \caption{(a) $R-V_g$ curve for the pristine sample (solid black line) and after exposure to light of $\lambda = 427$~nm $I=458~\mathrm{Wm^{-2}}$ for 3600s at $V_g^* = -5$~V. (b) Plot of the the change in normalized resistance with time on exposure to light exposure to light of $\lambda = 427$~nm.}
    \label{fig_s3}
   \end{figure}

In our heterostructure, there are two hBN flakes, one at the top and one at the bottom of the graphene. The discussion in the main manuscript assumes that the top hBN does not play any role in the doping process; note that all measurements reported in this work were performed with the back gate bias. To confirm this, we fabricated a device with only top hBN and illuminated it with the light of $\lambda=427$~nm at a back-gate voltage of $V_g=-5$~V. As shown in  Fig.~\ref{fig_s3}, the shift in Dirac point, even after prolonged exposure of one hour with high-intensity light, is very small, significantly less than the doping effect observed in our original device, which was $-2$~V in 50 ms. Hence, the contribution from the top hBN is negligible, and the bottom hBN is responsible for the doping effect.

\section{\textbf{Doping at low temperatures}}
\begin{figure}[h]
\centering
   \includegraphics[width=\columnwidth]{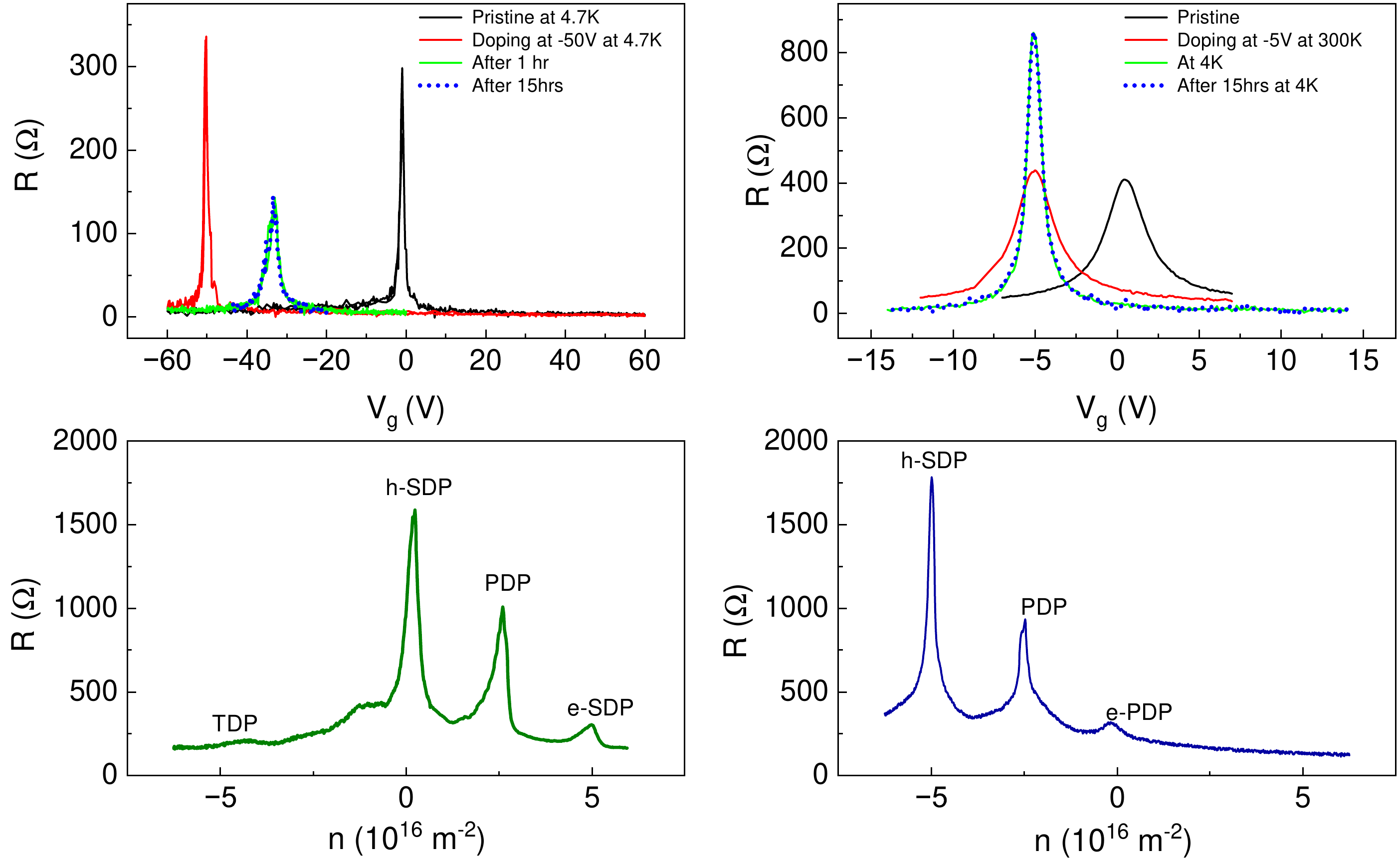}
  \caption{(a) $R-V_g$ curve for the pristine sample (solid black line) at room temperature and after doping at $V_g^* = -5$~V (solid red line) at room temperature. $R-V_g$ curve at $4.7$~K (solid green line) immediately after cool-down and after 15~hours (dash blue line). (b) $R-V_g$ curve for the pristine sample at 4.7K (solid black line); $R-V_g$ curve immediately after doping at $V_g=-50$~V (solid red line) and after 15 hours (dash blue line) of doping. (c) $R-V_g$ curve for electron-doped moire device with primary Dirac point (PDP) shifted to  $n=-2.5\times10^{16} \mathrm{m^{-2}}$. The hole-side tertiary Dirac point (TDP) is accessible along with the electron-side secondary Dirac point (e-SDP) and the hole-side secondary Dirac point (h-SDP). (d) $R-V_g$ curve for hole-doped moire device with primary Dirac point shifted to  $n=2.5\times10^{16} \mathrm{m^{-2}}$.}
    \label{fig_s4}
   \end{figure}

 Quantum transport measurements are performed at low temperatures. We tested the compatibility of our doping process with low-temperature measurements. The sample was doped at room temperature for the low-temperature experiments and then cooled down in a cryostat. As shown in Fig.~\ref{fig_s4}(a), doping was retained at low temperatures, and the $R-V_g$ response of the sample remained unchanged even after $15$ hours when kept cool. By contrast, when the doping was done at lower temperatures, we found the device did not retain the doping, as shown in Fig.~\ref{fig_s4}(b). Hence, the doping should be done at room temperature, and then the device can be cooled down.

An important application of the photodoping process discussed in this work is in accessing high-energy regions of the bands in graphene, which can not be reached by traditional electrostatic doping. Using this method,  we doped an hBN/SLG moir\'e device to electron and hole number densities of $n=2.5\times10^{16}~\mathrm{m^{-2}}$ as shown in Fig.~\ref{fig_s4}(c,d). We can access the tertiary Dirac point ($n=6.8\times10^{16}~\mathrm{m^{-2}}$) on the hole side; this was inaccessible in previous measurements due to gate-leakage during standard electrostatic gating.

\section{\textbf{Minimum detectable power}}

\begin{figure}[h]
\centering
   \includegraphics[width=0.5\columnwidth]{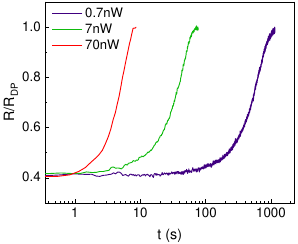}
  \caption{Plot of the change in normalized resistance with time for electron doping at $V_g=-3$~V for light of $\lambda=420$~nm and $P=0.7$~nW (solid blue line); $P=7$~nW (solid green line) and  $P=70$~nW (solid red line).}
    \label{fig_s5}
   \end{figure}

The device was illuminated with different powers of light of $\lambda = 410$~nm at negative gate voltage $V_g=-3$~V, and the corresponding resistance versus time curves were recorded as shown in figure Fig.~\ref{fig_s5}. The illumination power ($P$) was measured using a power meter. We could observe a change in the resistance for $P$ as low as $0.7$~nW.

\clearpage

\bibliography{arXiv}
\end{document}